\documentclass[a4paper]{article}

\usepackage{INTERSPEECH2019}
\usepackage[utf8]{inputenc}

\title{Attention model for articulatory features detection}
\name{Ievgen Karaulov$^1$, Dmytro Tkanov$^1$}
\address{
  $^1$Sciforce, Ukraine}
\email{ekaraulov@sciforce.solutions, dtkanov@sciforce.solutions}

\begin{document}

\maketitle
\begin{abstract}
% Abstract should be <=200 words long.
Articulatory distinctive features, as well as phonetic transcription, play important role in speech-related tasks: computer-assisted pronunciation training, text-to-speech conversion (TTS), studying speech production mechanisms, speech recognition for low-resourced languages.
End-to-end approaches to speech-related tasks got a lot of traction in recent years. We apply Listen, Attend and Spell~(LAS)~\cite{Chan-LAS2016} architecture to phones recognition on a small small training set, like TIMIT~\cite{TIMIT-1992}. Also, we introduce a novel decoding technique that allows to train manners and places of articulation detectors end-to-end using attention models. We also explore joint phones recognition and articulatory features detection in multitask learning setting.
\end{abstract}

\section{Introduction}

End-to-end approaches emerged in neural translation and later significantly changed automatic speech recognition (ASR) and TTS. While conventional pipelines still provide decent results, especially on smaller datasets, end-to-end models quickly catchup and are already state-of-the-art on some tasks~\cite{Zeyer-IS2018}. End-to-end models are typically sequence-to-sequence models that output words, subword units, graphemes or phonemes directly.

Articulatory features~\cite{Chomsky-Halle-1968} play an important role in describing speech production mechanisms. They can be divided in voicing, manners and places of articulation. A combination of a manner, a place and voicing defines any sound that human can produce in a unique way. For example, by convention ``s'' in the word ``sea'' is a voiceless alveolar sibilant which contrasts it to other phones.

By their nature, articulatory features are language agnostic and open many research oportunities in crosslingual and multilingual speech recognition, as linguistic features in speech synthesis etc. Particularly, Interspeech~2017 paper by Abraham~\cite{Abraham-IS2017} explores these features for ASR in low-resource setting. Automatic Speech Attribute Transcription (ASAT) system~\cite{ASAT-2007},~\cite{ASAT-2014} also works on producing speech-to-text model based on indicators detection (subset of these indicators are articulatory features; hereafter both terms are used with the same meaning).

Another application of phones recognition and articulatory features estimation is Computer-Assisted Pronunciation Training. Some of the approaches are described in~\cite{Eskenazi-2009} and~\cite{Ryu2017}.

This paper dwells on applications of attention-based models to articulatory features detection. We introduce a novel decoding technique that allows training manners and places of articulation detectors end-to-end with the help of attention-based models. We also explore joint phones recognition and articulatory features detection in a multitask learning setting. Contrary to other works in this domain, we focus on producing sequences of features instead of frame- or segment-level labels. Sequence-level data is simpler to work with in applications where precise alignment with original waveform is not important. It also may serve as input for the encoder in sequence-to-sequence-based speech synthesis. Besides, articulatory features are language-independent, and our approach can potentially be applied to zero-resource speech recognition.

\section{Previous work}
The conventional approach to estimation of phonological features is akin to the standard ASR pipeline. It requires forced alignment of phones to utterances. As a result, training is usually done either on fine-labeled data with alignments or on data that have good acoustic models available. This limits research to well-studied mainstream languages or enforces usage of cross-language models.

Reasearch on detection of places and manners of articulation is spread across different languages and it is challenging to select a single baseline. We decided to use TIMIT in our experiments as it is an overall well-studied corpus with well-known published results for phones recognition~\cite{Graves2013} and articulatory features detection~\cite{King-2000}.

While we focus on detection, there is still a noteworthy series of recent works on manners discrimination using zero-time windowing, e.g.~\cite{Prasad-2018}, which might be used in feature engineering for encoder inputs.

For non-English baselines, a good starting point is work by Merkx and Scharenborg~\cite{Merkx-IS2018} that describes positive impact of CNNs on ariculatory features classification in Dutch.

Another end-to-end approach to articulatory features detection is described in~\cite{Qu-2018}, where authors focus on connectionist temporal networks (CTC)~\cite{Graves2006}.

\section{Model description}
\subsection{Attention-based models}
Typical end-to-end models in speech domain are based on sequence to sequence neural networks. Most common architectures are CTC, recurrent neural network transducer (RNN-T)~\cite{Graves2013} and encoder-decoder with attention~\cite{Chorowski-Att-2015}. In this work we will focus on attention-based models. One of important features of attention-based models is that they provide a link between the encoder (acoustic data) and the decoder (textual data or indicators) steps. For training this link results in faster convergence to lower loss values. Besides, during inference, attentions enables building a distribution that can relate the sound data and the textual data, essentially providing fuzzy alignment between the decoder outputs and the encoder inputs.

\subsection{Articulatory features}
Articulatory features describe production of sounds via the interaction of different components within a human vocal tract.
Our approach rests largely on the seminal paper by Simon King~\cite{King-2000} that used so-called Spoken Patterns of English (SPE~\cite{Chomsky-Halle-1968}) features and manners as targets instead of phones for training a neural network.

We used a combination of SPE, manners and places of articulation in our experiments.

\subsection{Multi-task learning}
Multi-task learning (MTL)~\cite{Caruana98} improves learning efficiency and model generalization for the task-specific models by learning several related tasks at the same time. All tasks usually share a part of representation. Each new task contributes to the model learning by adding information and transferring knowledge. MTL approach is applied to neural networks by sharing some of the hidden layers between different tasks.

In our work, we share encoder parameters between two tasks: articulatory features classification and phonemes classification.

\subsection{Decoder for indicators}
In this work we train a LAS model on phone and indicator targets. For phone targets, the setup is rather straightforward: decoder outputs ARPABET symbols directly at each step. We do not use any language model, so the decoder outputs are passed for the error rate calculation as is.

For indicators detection, we modified the decoder. We use a sigmoid activation function for the projection layer, since these features are not mutually exclusive. At each step, the decoder outputs posteriograms for places and manners of articulation. Our approach has two options for the indicators decoder. The first option outputs samples from posteriors of indicators. Alternatively, the indicators posteriors are transformed to phones posteriors via the mapping matrix $ M=\{m_{ij}\} $, $i=1..M$, $j=1..N$, where $ m_{ij} $ is a binary indicator of presence or absence of the corresponding feature $ f_i \in F $ in the phone $ s_j \in S $; $ M $ is a phones count in the lexicon, $ N $ is an indicators count.

Let's describe the second approach in detail. For clarity, we are describing all the operations during inference. Let $ \Phi $ be a set of posterior probabilities $ \phi_i $ of all indicators at encoder's output and decoder's time step $ t $. Assuming binary features' independence, we can obtain log probabilities of observing phones $ P=\{\log p_i\} $ using the following equation:
\begin{equation}
	P = \log{\Phi} \cdot M + \log{(1 - \Phi)} \cdot (1 - M)
\end{equation}

After obtaining log posteriors of phones $ P $, we then find one with the highest probability $\hat{f}=f_{\hat{j}}$, $\hat{j} = \textrm{argmax} P $, and perform its reverse mapping on articulatory features vector by just getting column $ \hat{j}$ from $ M $. This new vector of refined indicators is passed to the decoder's input for calculating output at decoder's next time step $ t + 1 $. Thus, binary features posteriors are rounded to the values corresponding to the closest phone, filtering out invalid combinations which never occur in the language (for example, front and back, stop and continuant, etc.).

It is important to note that the set $ S $ of all possible phones is not constrained by phonetic inventory of training language and thus can be used in cross-lingual low resource speech recognition. So, by specifying a mapping matrix for a different language than used at training time, we can perform cross-lingual phones recognition.

Finally, we apply MTL for both phone targets and indicators. Both tasks share the encoder, however, the attention and the decoder are distinct for them. Losses of both tasks contribute equally to the overall loss on each step. In our experiments, the joint model needs 20\% less steps for convergence than separate models. A possible explanation is that a loss for indicators targets of similar phones (e.g. target ``n'' and network output ``ng'') will have lower value than a loss for distant phones (e.g. ``n'' and ``ow''), due to matching articulatory features, even at early training steps. At the same time, for phone targets, network outputs that are closer to targets will have a higher loss, at least at early stages of training.

\section{Experiments}
\subsection{Dataset and features}
All experiments were performed on the TIMIT corpus. For training we used the standard 462-speaker set without SA records. Test results in tables below correspond to the core test set of 192 utterances. A development set was collected from the remaining part of the test set, i.e. the non-core part. We explicitly checked that the train, development and test sets had non-overlapping speakers. Also, the test core set does not share either speakers, or phrases with the train and development sets.

There are 61 phone types in the corpus, yet a common practice is to report results on a reduced 39-phone set with allophones mapped to a common phonetic label following~\cite{Lee-1989}. Also most papers perform training on the full phone set and do 61 to 39 mapping for evaluation. In our experiments we did not find any benefits of this approach against direct training on the reduced phones set. Both ways (the full set with mapping and the plain reduced set) yielded near identical results.

For features extraction we tried several approaches. We followed~\cite{Graves2013}: 20~ms window and 10~ms step mel-scale filterbanks plus log-energy feature. Also we experimented with mel-frequency cepstral coefficients (MFCC) and Lyon's auditory model~\cite{Lyon-1982}. In each case of base acoustic features we computed deltas and double-deltas. All inputs were globally normalized to have zero mean and unit variance for each input feature.

\begin{table}[th]
  \caption{Phone error rate (PER) comparison for different feature sets on TIMIT corpus.}
  \label{tab:FeaturesTimit}
\centering
  \begin{tabular}{ll}
    \toprule
    \textbf{Feature} & \textbf{PER} \\
    \midrule
    Melfilterbanks & 21.7\% \\
    MFCC & 21.8\% \\
    Lyon & 21.2\% \\
    \bottomrule
  \end{tabular}
\end{table}

Table~\ref{tab:FeaturesTimit} compares three models with around 6.5 million weights and identical architectures: 3 layers of 256 units in encoder, 1 layer of 256 units in decoder, Luong attention, 20\% dropout and 10\% scheduled sampling probabilities. We do not do any forced alignment and do not use temporal alignments for phones provided with TIMIT, thus phone error rates (PER) reported in this paper are essentially edit distances normalized by ground truth length. After conducting experiments we concluded that with other parameters fixed, MFCC and melfilterbanks perform similar in terms of phone error rate while Lyon's cochleogram provides a slightly better accuracy. Still, to make comparison with other baselines more fair, we conducted further training on MFCC features.

\subsection{Training}
We performed training on a single NVIDIA~GTX1080~GPU. On average, one training run takes up 3-5~hours depending on model size. We tried to keep a batch size of 32, yet occasionally had to lower it for some experiments to 16 due to memory limitations. In all experiments we used a decoder with a single layer. As to the encoder-decoder connection, we noticed that attention provides enough information to the decoder and passing the final encoder state to the decoder does not yield any performance improvements.

TIMIT is a small dataset with a few hours of speech in all sets combined. Thus, deep learning models overfit easily and regularization plays a crucial role. We relied on dropout and L2 weight decay for regularization. We used drop probability in 20 -- 40\% range and decay constant between $10^{-5}$ and $10^{-3}$.

\subsection{Results}
In our experiments, we did not use external language models or speaker adaptation. We used LSTMs in the encoder without any initial convolution layers. Adding any of this improvements should positively impact accuracy of the resulting model.

\subsubsection{Phone recognition}
We used the results from~\cite{Graves2013} and~\cite{Chorowski-Att-2015} as baselines to compare our results against them. Their phone error rates for pretrained transducer and attention-based models are among the lowest reported for sequence-to-sequence models that do not rely on external language models.
\begin{table}[th]
  \caption{Phones recognition on TIMIT corpus. LAS -- a model trained on phone targets. LAS-F -- a model trained on articulatory features targets. LAS-MTL-S -- a multitask model with sampling from the decoder outputs. LAS-MTL-M --a  multitask model with the decoder inputs converted using a mapping matrix.}
  \label{tab:PhoneRecTimit}
\centering
  \begin{tabular}{llll}
    \toprule
    \textbf{Paper} & \textbf{Model} & \textbf{Parameters} & \textbf{PER} \\
    \midrule
    Baseline~\cite{Graves2013} & Transducer & 4.3M & 17.7\% \\
    Baseline~\cite{Chorowski-Att-2015} & ARSG & around 6M & 18.7\% \\
    Ours & LAS & 5.6M & 20.2\% \\
    Ours & LAS-F & 5.6M & 23.4\% \\
    Ours & LAS-MTL-S & 7.1M & 20.4\% \\
    Ours & LAS-MTL-M & 7.1M & 20.8\% \\
    \bottomrule
  \end{tabular}
\end{table}

Table~\ref{tab:PhoneRecTimit} shows the comparison of our models against baselines. At first glance, our implementation of a multitask attention-based model (LAS-MTL-S and LAS-MTL-M) for joint articulatory features and phones recognition performs worse than baselines. We attribute this to a suboptimal choice of hyperparameters, because the non-multitask model (LAS) trained on just phone recognition yields results comparable to the multitask model. Therefore, adding indicators as an additional target doesn't decrease the phone recognition accuracy. Another interesting observation is that the model LAS-F trained directly on articulatory features without complementary phone targets results in a higher error rate than LAS-MTL models. In this case, PER serves as integral accuracy for all indicators together. A review of individual features accuracy also showed that the multitask learning approach yields higher accuracy than just using articulatory features targets.

Besides, our experiments show that sampling from indicators posteriors (LAS-MTL-S) performs better than using a mapping matrix (LAS-MTL-M). A possible explanation is that articulatory features can express minor variations in phones pronunciation (such as assimilation or coarticulation) that are not reflected in the phonetic alphabet.

An important takeaway from this experiment is that a sequence-to-sequence articulatory features detection model, besides actual speech indicators, also yields phone recognition results close to strong baselines.

\begin{figure}[ht]
  \centering
  \includegraphics[width=\linewidth]{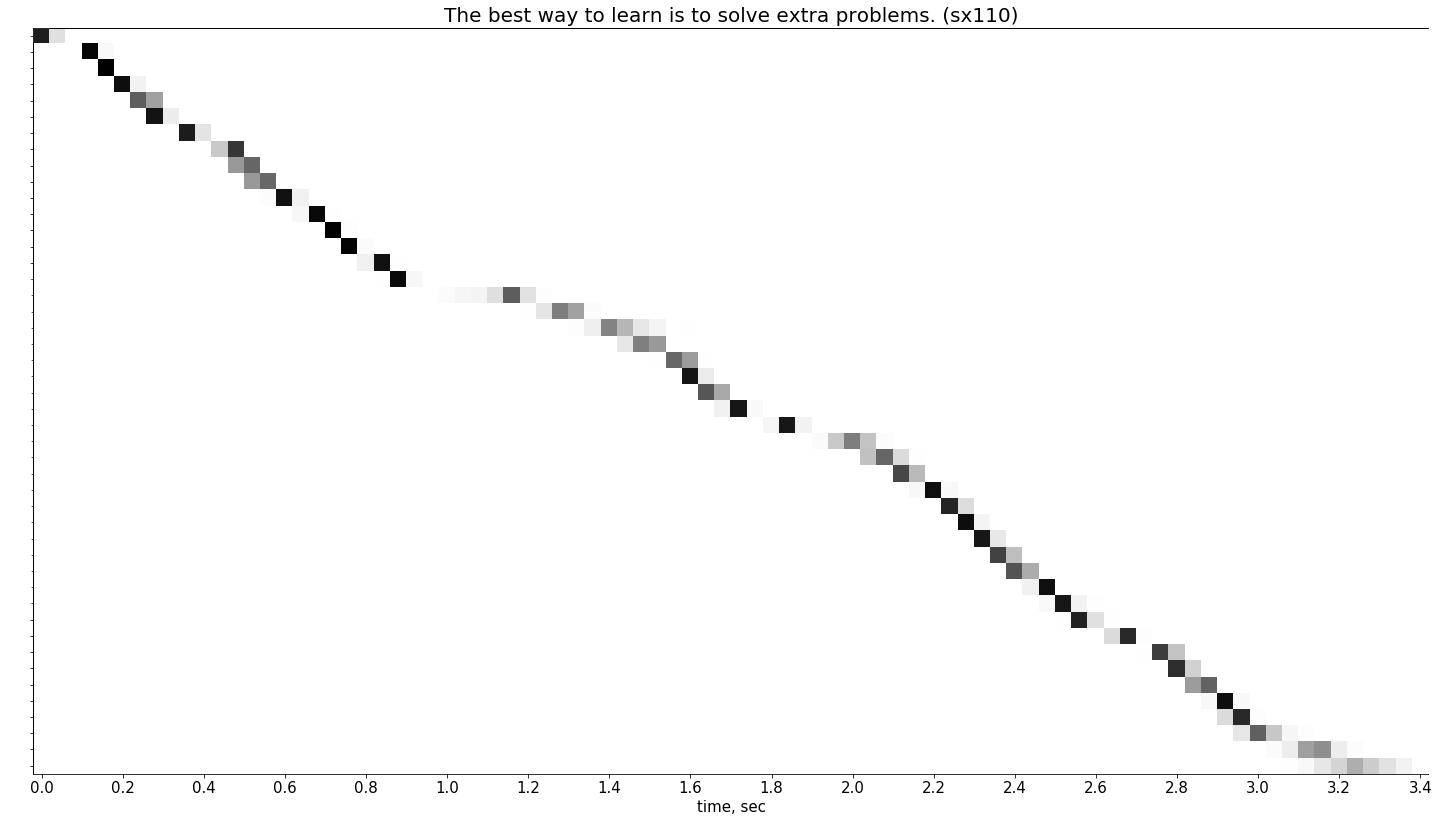}
  \caption{Attention output for phrase SX100 from test set: ``The best way to learn is to solve extra problems''.}
  \label{fig:alignment}
\end{figure}

Figure~\ref{fig:alignment} shows the output of attention for phrase SX100 ``The best way to learn is to solve extra problems'' from the test set. The model learned to align indicators with acoustic data. Mapping articulatory features to phones yields the following transcription: [sil dh ah sil b ae s sil t w ey sil t ah dh er n ih z sil t ih s aa l v eh sil k s sil t er sil p r aa sil p l m s sil].

\subsubsection{Manners and places of articulation detection}

\begin{figure}[ht]
  \centering
  \includegraphics[width=\linewidth]{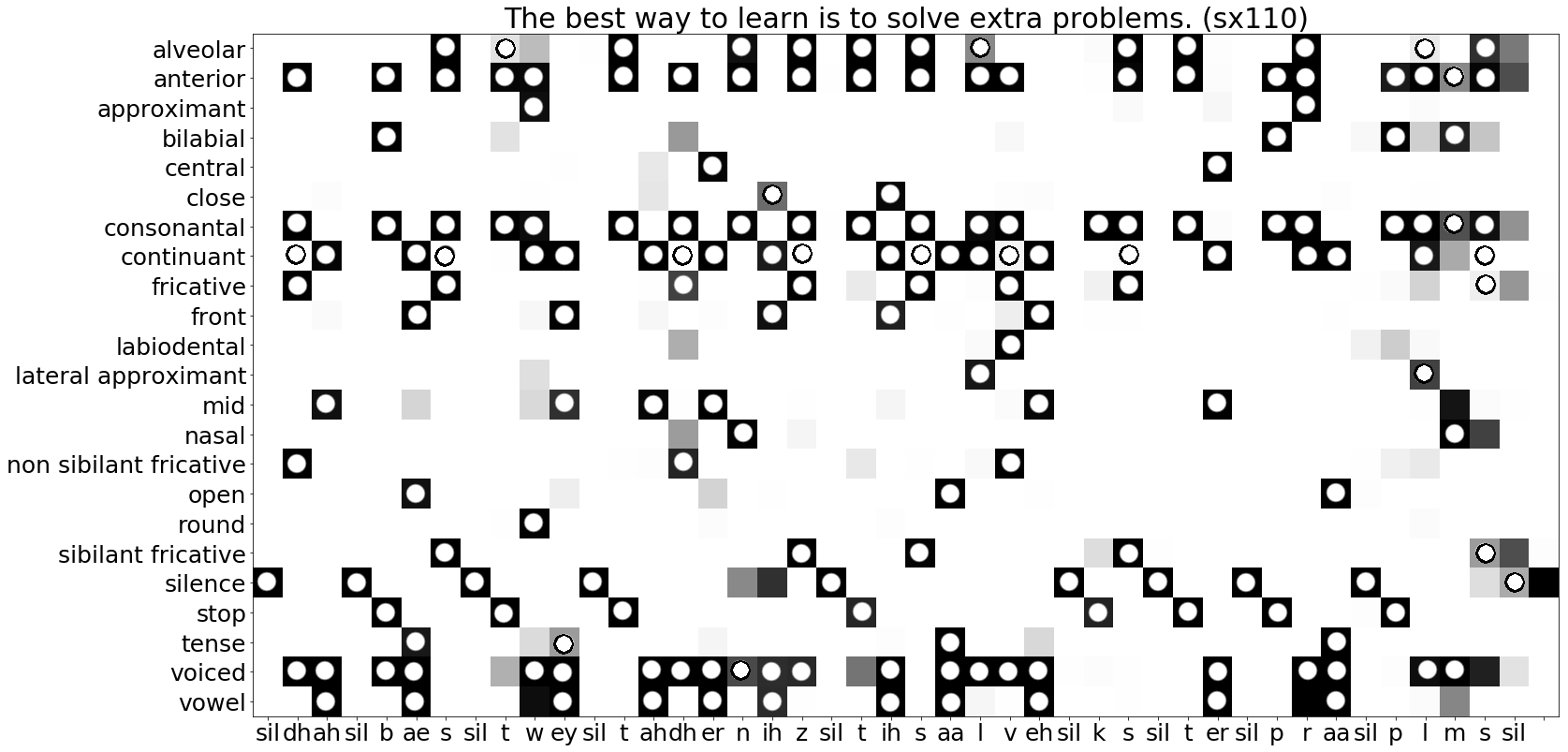}
  \caption{Articulatory features posteriors for phrase SX100 from the test set: ``The best way to learn is to solve extra problems''. Ground truth features are marked with white circles. }
  \label{fig:binf}
\end{figure}

In figure~\ref{fig:binf} we can see outputs of the multitask model decoder for indicators on SX100 utterance. These are raw values without ``rounding'' to the closest phones.

To the best of our knowledge, there is no previously published results for sequence-level articulatory features detection on TIMIT. In~\cite{Qu-2018} the authors report articulatory features detection results on Wall Street Journal (WSJ) corpus~\cite{WSJ}. The accuracy is in 91--96\% range, yet to do a proper comparison, we might need to train a model on WSJ.

Having no other options, we compare to frame-level baselines available. For King and Taylor paper~\cite{King-2000}, we provide results from experiment 2 which has a broader features set that is closer to what we used. ASAT~\cite{ASAT-2007} paper mentions only a few accuracy values and some of them are suboptimal, for example, 75\% for close and 68\% for mid places. The reported phone error rate for ASAT is 29.4\%.

Using alignments from attention, it is possible to map sequence-level posteriors to acoustic frames from the original waveform. It is important to understand that a direct projection of sequence outputs to frames is not precise. Still, we provided accuracy of a frame-level projection to show that even without direct training on frame targets, alignment makes sense in spite of pyramidal stacks of recurrent layers in the encoder.
\begin{table}[th]
  \caption{Manner and places of articulation on Timit corpus. Ours -- LAS-MTL-M model, KT -- King and Taylor paper~\cite{King-2000}. Accuracy rates are reported at frame and sequence levels. For frames level, our model accuracy is computed with a direct projection to frames and with a projection to markup segments.}
  \label{tab:BinfTimit}
\centering
  \begin{tabular}{llll|l}
    \toprule
                     &  & \textbf{Ours} & & \textbf{KT} \\
    \midrule
                     & symbols & markup & & \\
    \textbf{Feature} & sequence & frames & frames & frames \\
    \midrule
alveolar & 94\% & 95\% & 77\% \\
anterior & 94\% & 90\% & 69\% & 90\% \\
approximant & 98\% & 98\% & 94\% & 68\% \\
bilabial & 98\% & 98\% & 93\% \\
central & 98\% & 99\% & 91\% \\
close & 96\% & 97\% & 88\% & 86\% \\
consonantal & 94\% & 88\% & 64\% & 90\% \\
continuant & 97\% & 89\% & 68\% & 86\% \\
fricative & 96\% & 95\% & 83\% & 88\% \\
front & 96\% & 95\% & 89\% & 84\% \\
glottal & 99\% & 99\% & 98\% \\
labiodental & 99\% & 99\% & 96\% \\
lateral approx. & 98\% & 99\% & 96\% \\
mid & 93\% & 97\% & 82\% & \\
nasal & 98\% & 99\% & 93\% & 84\% \\
non-sibilant fric. & 97\% & 97\% & 94\% \\
open & 98\% & 98\% & 91\% & 93\% \\
palatal & 99\% & 99\% & 99\% \\
postalveolar & 99\% & 99\% & 97\% \\
round & 98\% & 98\% & 91\% & 92\% \\
sibilant affric. & 99\% & 99\% & 99\% \\
sibilant fric. & 98\% & 98\% & 90\% \\
silence & 97\% & 80\% & 63\% & 89\% \\
stop & 97\% & 97\% & 85\% & 96\% \\
tense & 96\% & 97\% & 81\% & 87\% \\
velar & 99\% & 99\% & 95\% \\
voiced & 95\% & 84\% & 72\% & 93\% \\
vowel & 97\% & 92\% & 70\% & 92\% \\
    \midrule
PER & 20.8\% & 24.8\% &  & 40\% \\
    \bottomrule
  \end{tabular}
\end{table}

Table~\ref{tab:BinfTimit} shows detection accuracy for manners and places of articulation.
Our attention-based model for articulatory features provides accurate sequence-level estimates with most manners and places being in 95\%+ range. PER and indicators accuracy imply that even if the decoder yields a wrong phone, most of articulatory features inferred would align with ground truth.

To calculate frame accuracies presented in the ``Ours frames'' column, we used the following phone-to-frame mapping algorithm. Soft attention was converted to hard attention using the dynamic time warping algorithm (DTW)~\cite{Salvador-2007}. Ambiguities in phone-to-frame mapping were resolved by leaving points with higher attention weights on the alignment path.

Inaccuracies on the frame level (``Ours frames'' column) are almost exclusively caused by boundary frames: attention outputs a peak at the prominent acoustic part of a phone sound and the projection for the remaining part becomes ambiguous. This effect is especially clear for long phones, e.g. vowels, and for long silences where a single ``sil'' symbol is mapped to 1 or 2 starting frames from the final encoder layer.

It is possible to do a more precise projection by using attention peaks as a phone nuclei estimator and combining them with other waveform segmentation methods. In this manner, the frame-level accuracy would be much closer to the sequence level. To illustrate this point, we projected predicted indicators to waveform segments from TIMIT markup using attention peaks (the ``Ours markup frames'' column in table~\ref{tab:BinfTimit}). As a result, accuracy for all features increased significantly. For some features, e.g. vowels, accuracy went up by 20 percent points. Theregore, a combination of a LAS-based model with an external waveform segmentation model may lead to more precise alignments of detected features with the acoustic signal.

\section{Discussion}
A LAS encoder is a stack of pyramidal layers. As a result, the top-most layer has a typical window step of 40 -- 80 ms depending on the number of encoders layers. This interval determines inaccuracy that would be even in case of the ideal projection of sequence symbols to frames through attention. One way to infer more accurate phone boundaries within 80ms superframes is to use conventional methods for speech segmentation.

An interesting side note is that independent from input features or model hyperparameters, a single most common mistake made by our models was ``ah'' -- ``ih'' substitution. It accounted for more than 0.5 percent point of all mistakes.

It is worth noting that TIMIT has explicit phone sequences for each utterance, yet it is possible to perform training on corpuses with just textual transcript by applying grapheme-to-phoneme conversion. One way to do this is to use the ``espeak-ng\footnote{https://github.com/espeak-ng/espeak-ng}'' software that has various models for conversion from graphemes to International Phonetic Alphabet. As a result, it is possible even to train a multilingual model on several corpuses for different languages.

\section{Conclusions}
The paper proposes a novel approach to end-to-end articulatory features detection. The resulting model yields posteriorgrams for articulatory features, rough alignments with acoustic data and competitive phone error rates even in low-resource settings.

In future, we would like to study the possibilities of applying our approach to recognition of phones and articulatory features on a zero-resource language using the model trained on a higher-resource language(s). Besides, we would like to study more carefully the ways to improve time alignment of the model outputs with the waveform.

The code we used to train and evaluate our model is available at \textit{https://github.com/sciforce/phones-las} .

\section{Acknowledgments}
We would like to thank Tzu-Wei Sung from National Taiwan University for his implementation of ``Listen, Attend and Spell'' model\footnote{https://github.com/WindQAQ/listen-attend-and-spell} that we used as a starting point for our experiments. Also we would like to thank Olga Zvyeryeva for preparing mappings from phones to articulatory features.

\bibliographystyle{IEEEtran}

\end{document}